\documentclass[%
 aip,
rsi,%
 amsmath,amssymb,
 reprint,%
]{revtex4-1}

\usepackage{graphicx}
\usepackage{dcolumn}
\usepackage{bm}

\begin{document}

\preprint{AIP/123-QED}

\title{Quantitative analysis of numerically focused red blood cells using subdivided two-beam interference (STBI) based lateral-shearing digital holographic microscope}

\author{Shital Devinder}
\affiliation{Instrument Design and Development Center, IIT Delhi}
\author{Ashish Lal}%
\author{Tathagato Rai Dastidar}
\affiliation{%
SigTuple Technologies Pvt Ltd. Bangalore, India.
}%
\author{Satish Kumar Dubey*}
\altaffiliation{Instrument Design and Development Center, IIT Delhi}
\homepage{http://web.iitd.ac.in/~satishdubey/}
\email{satishdubey@iddc.iitd.ac.in}
\date{\today}

\begin{abstract}
A lateral shearing interferometer based digital holographic microscopy has been realized to study the morphology and dynamics of the biological samples quantitatively. Here a lateral shearing interferometer is embedded with a conventional microscope and a CMOS sensor to form a lateral shearing digital holographic microscope. It enables recording of the image plane holograms of the object that can be numerically reconstructed to estimate its 3-D profile. This yield the depth information of the sample in addition to its bright field image. To overcome the duplicate image problem, a method called subdivided two-beam interference (STBI) has been used that enables the recording of the hologram equivalent to the two-beam holographic microscope. Since a small region of the smear is sufficient to study its cellular morphology. Here region of interest is isolated from the rest of the sample and STBI method is applied to analyze the peripheral blood smear quantitatively. In addition, focusing on the cells is ensured using a computational approach called total variation (TV) method. It is planned to automate the process to get the best focused 3D profile of the object (cells). In this study, STBI and TV approaches were applied to record and reconstruct the holograms corresponding to the human blood smear.
\end{abstract}

\maketitle

\section{\label{sec:level1}Introduction:}

Digital holographic microscopy using two beam configuration is a common method used for the phase imaging of microscopic objects like RBCs, WBCs, platelets etc~\cite{6255748} ~\cite{article1}~\cite{Mann:05}. Phase and amplitude information of the object are encoded on a charged couple device(CCD) or on a complementary metal-oxide-semiconductor (CMOS) camera in the form of interference fringes called the hologram. Phase and amplitude information is retrieved from the digitally recorded holograms using numerical reconstruction methods ~\cite{Schnars_2002}.Two beam configuration enables recording of wide field of view (FOV) of the sample under investigation. Also, it is possible to have wider angular tilt between object and reference beams. The angular tilt need to be higher than a certain level to avoid the overlap of dc and cross terms in the Fourier domain. However there are techniques for high quality image recovery even when the dc and the cross terms in the hologram overlap in the Fourier domain.~\cite{Khare:13} ~\cite{Zhang:04}. One such configuration is Machzender Interferometer based digital holographic microscope (DHM). In this configuration, a coherent beam is split into two, one illuminates the object and other is kept unaltered. These wavefronts follow different paths in space and are so directed to produce interference fringes on CCD. The separation of beams in two arms of setup leads to the reduced temporal phase stability which results in the fluctuations of phase of object over time. This types of configuration also requires more number of optical components, making the setup difficult to align, bulky and expensive. This problem has been addressed using a common path configuration called lateral shearing digital holographic microscope(LS-DHM)~\cite{Singh:12}~\cite{article}~\cite{ Singh:18} having a high temporal phase stability. This configuration uses almost half of the optical components that are used in standard two beam interferometer based DHM making it easier to align, compact and cost effective. However this configuration suffers from duplicate image problem which emerges due to the self referencing geometry. Self overlap of the object beam also leads to the loss of the object information. Object information remains available only in the region where there is a overlap of altered region (having object information) with that of unaltered (no object information) region of sheared wavefront. The problems of object information overlap and the duplicate image have been addressed in the present manuscript. In this regard, following two  approaches have been incorporated, (i) scanning the peripheral region of the sample under investigation and (ii) maintaining the lateral shear distance approximately equal to the half of the field of view. Moreover, accurate focusing of the field of view is essential for the quantitative analysis of the blood smear because even a small defocus alters the phase of the object considerably. The defocus may result in the alteration of the features like cell volume, surface area etc. by great deal which is used for the identification and classification of different type of cellular objects.  In two beam interference based DHM, sample is focused by blocking the reference beam and recording a bright field image of highest contrast. Unblocking the reference beam results in a well focused hologram. However, this is not possible in LS-DHM, as interfering beams are always overlapping. In this manuscript, a computational approach has been proposed for finding the best focus hologram out of a stack of holograms recorded around the plane of the best focus.The best focused hologram is decided on the basis of the total variation (TV) method. Hologram with the highest TV value is considered to be the best focused hologram.

\begin{figure}[ht!]
\centering\includegraphics[width=6cm,height=8cm]{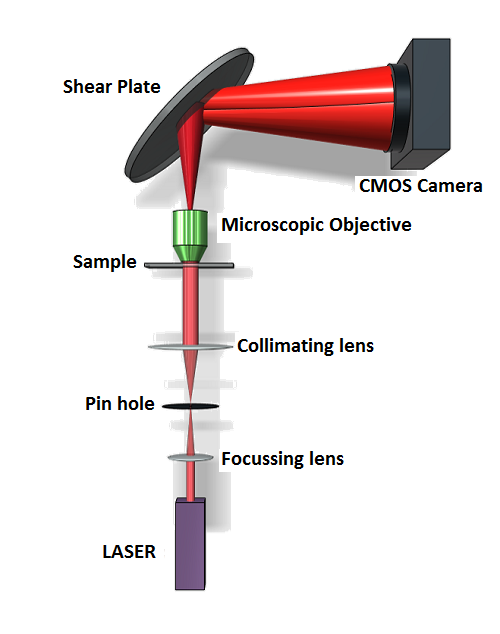}
\caption{Schematic of LS-DHM used for the study of blood cells.}
\end{figure}

\section{Methods:}
\subsection{Lab setup:}
The LS-DHM setup consists of a coherent light source, a microscopic objective for magnification, a wedged glass plate and a detector (CCD/CMOS camera). The schematic of the proposed setup is as shown in Fig.1. The used light source is a He-Ne laser ($\lambda = 633 nm, 5 mW, unpolarised$). Beam is spatially filtered using a $40X$ M.O. and a $10 um$ pinhole. A $100 mm$ focal length lens is used for collimation . The collimated beam is directed on the sample which is subsequently  magnified by a ($10X$) microscopic objective. This magnified version of the object wavefront is made to fall on a thick ($10 mm$) shear plate. The object beam is reflected from front and back surfaces of the shear plate. These two reflected object beams make an interference pattern, which is recorded on a CCD/CMOS camera. \\

\subsection{Subdivided two beam interference:} 
As a consequence of the conventional lateral shearing interferometer system, recorded hologram contains duplicate images and information overlap due to the interference between two sheared object beams. This overlap degrades the reconstruction results and makes it difficult to extract the phase of the object. To avoid overlap of object information, periphery of the sample is scanned which essentially yields the object wavefront as two sub-divided beams~\cite{article} with and without object information as shown in Fig.(2). 

\begin{figure}[ht!]
\centering\includegraphics[width=7.5cm,height=5cm]{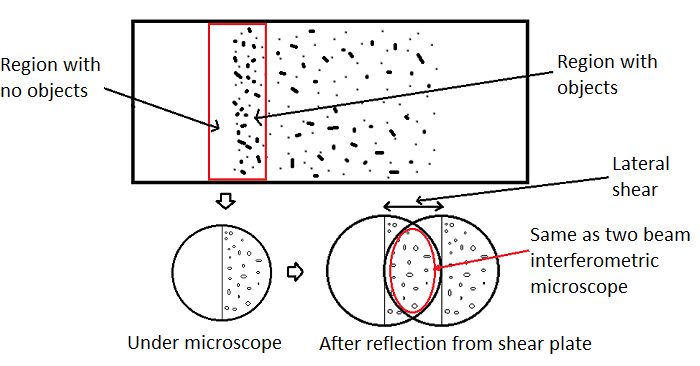}
\caption{Diagram showing STBI at the periphery of the blood smear.}
\end{figure}

In such a case if the lateral shear distance is made equal to half of the FOV, we get a hologram which is equivalent to that of the two beam interference DHM. Assuming the interference between altered part and the unaltered part of the object wavefront only, the recorded fringe pattern can be mathematically expressed as 
\begin{eqnarray}
        I_{recorded}(x,y)=|O_1(x,y)+O_2(x,y)|^2 
\end{eqnarray}
\begin{eqnarray}
    I_{recorded}(x,y)=O_1(x,y)^2+O_2(x,y)^2+O_1^*(x,y)\nonumber\\
                 O_2(x,y)+O_1(x,y)O_2^*(x,y)
\end{eqnarray}

where $O_{1}(x,y)$ refers to the altered part of the object wavefront, reflected from the front surface of the shear plate and $O_{2}(x,y)$ refers to the unaltered part of the object wavefront, reflected from the back surface of the shear plate. $O_{2}(x,y)$ is reflected from the back surface of the shear plate. It travels additional optical path, which adds a universal phase to the wavefront. This unwanted phase is removed by subtracting the background phase from the phase calculated when the object is present in the FOV. The subtraction leaves only the object phase.
\subsubsection{Focusing:}
For the extraction of accurate phase of microscopic samples, focusing of the sample is crucial. A small defocus leads to a significant phase error. In two beam interference microscope, sample can be brought to focus by blocking one of the beams (reference) and looking for the highest contrast image. The plane of highest contrast can be considered as the plane of focus. However, two beams always overlap in shearing interference microscope. It is difficult to first isolate and then look for the sharp bright field image. We propose a method to find the focused hologram out of a number of recorded holograms. As phase information of the object is encoded in terms of the modulated fringes, the maximum modulation of fringes correspond to the phase of the object. Therefore, a well focused hologram is one where in fringe modulations are maximum. Modulation in fringes can be calculated in terms of the total gradient of the image. Which is called as the total variation (TV) of the image. Total variation is the gradient of the image summed over all the pixels. Therefore, for the best focused hologram, total variation is maximum. Stack of holograms were recorded across the focus plane and one with the maximum TV was labeled as focused hologram.\\

\section{Recording a hologram}
 For recording  a hologram using the proposed system, region of interest must be chosen very carefully, taking into account the object location, thickness and refractive index of the shear plate and the angle of incidence at the shear plate. In the proposed microscope, hologram is produced by interference of two sheared object beams which are reflected from the front and back surfaces of the shear plate. In conventional systems, the duplicate image of the object could not be removed due to small lateral shear distance. This is because more than half of the object beam overlaps with itself. This problem is taken care by maintaining the lateral shearing distance less than or equal to half of the FOV in the proposed setup.
 
\section{Experiments:}
\subsubsection{Imaging of human red blood cells:}
In certain cases, diseased RBCs can be differentiated from healthy ones on the basis of phase. Phase of RBCs present in human blood sample is imaged using the proposed setup. The unstained blood smear was prepared in the lab and the periphery of the blood smear was scanned using a $10X$ M.O. (for wide FOV). The plate was kept at $25 cm$ from the microscopic objective at an angle of $45 degrees$ (approx.) to the object beam. The camera was kept at $20 cm$ from the shear plate. A raw video of holographic view was recorded on a $24 bit$ CMOS camera (frame rate $150 f/s$, pixel size of $4.8 um$) for approximately $10s$ while the sample was gradually moved in and out of the focus. Out of the large no of recorded frames, the frame with maximum TV value was considered for reconstruction and phase extraction. The feasibility of the TV method has been established to find out the best focused hologram. The procedure was repeated multiple times and the TV approach was found to be stable for the estimation of the focus position.\\

\subsubsection{Reconstruction:} 
In 20 seconds more than $3000$ frames were recorded. Total variation was calculated for $1^{st}$,$30^{th}$,$60^{th}$ frames and so on and the hologram frame with highest TV was considered for reconstruction. The numerical reconstruction of the hologram was carried out using Fourier transform method with accurate carrier removal technique~\cite{Singh:16}. The retrieved phase was unwrapped using the transport of intensity method~\cite{Pandey2018}.

\begin{figure}[ht!]
\centering\includegraphics[width=8cm,height=6.5cm]{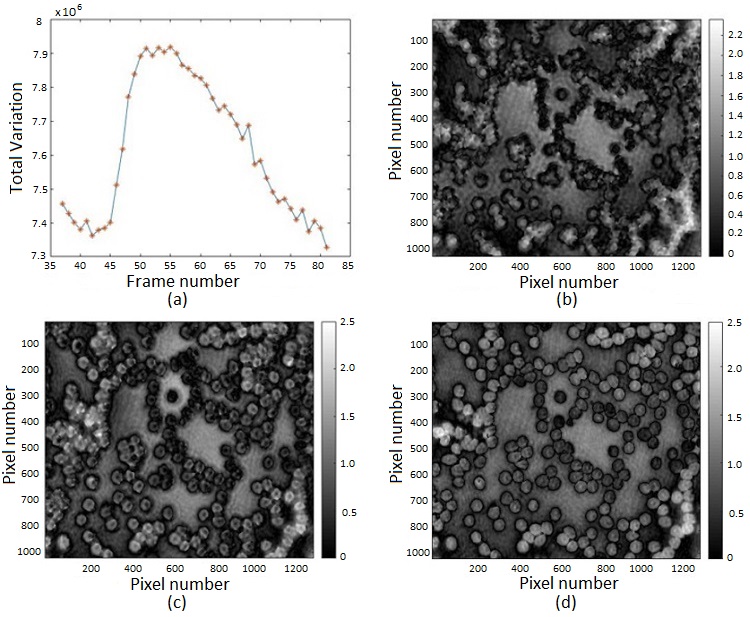}
\caption{(a) TV plot for various holograms recorded along microscope axis. (b) Phase map for frame no. 40. (c) Phase map for frame no. 58. (d) Phase map for frame no. 50. }
\end{figure}

\begin{figure}[ht!]
\centering\includegraphics[width=8cm,height=6cm]{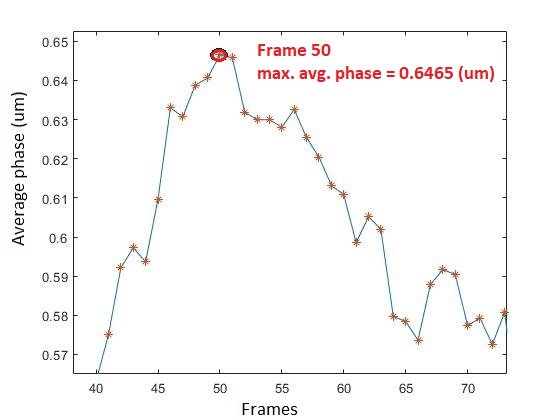}
\caption{Mean phase variation for stack of 30 holograms in the neighbourhood of the focused one }
\end{figure}

\begin{figure}[ht!]
\centering\includegraphics[width=8cm,height=6cm]{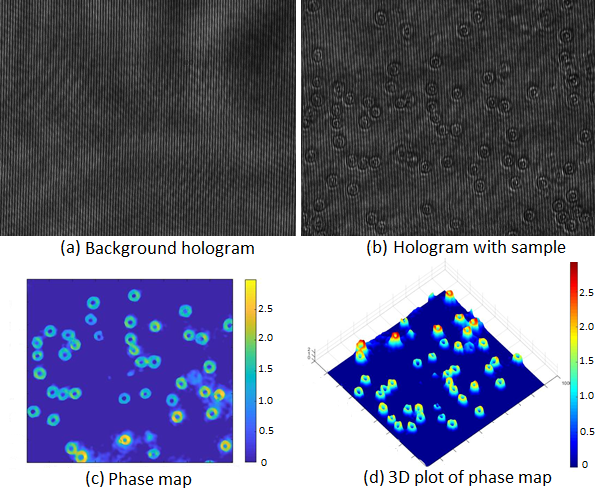}
\caption{(a) Background (Hologram recorded with no sample present). (b) Hologram recorded for the human blood smear. (c) Phase map for the recorded hologram. (d) Phase map 3D view of the recorded hologram.}
\end{figure}

\section{Results:}
\textbf{Focusing:} $Fig.3 (a)$ shows the variation of TV for the various frames recorded along the optical axis. X-axis stands for the $n^th$ no. of hologram and Y-axis shows the TV values for corresponding holograms. $Fig.3 (b)$,$fig.3 (c)$ and $fig.3 (d)$ show the phase map for the $40^{th}$, $58^{th}$ and  $50^{th}$ frame. The average phase for these frames is $0.5632$ , $0.6204$ and $0.6465$ respectively. $Fig. (4)$ Shows the average phase variation in the vicinity of focus. When in defocus, whether in positive or negative direction, the average phase is relatively smaller. Average phase is the maximum for the best focused hologram.\\

\textbf{Phase results for Blood cells} $Fig.5 (a)$ shows background hologram i.e. without sample being present and $fig.5 (b)$ hologram with sample (blood smear). Holograms for RBCs are recorded at the peripheral smear without information overlap. Moreover RBCs at the boundary are least effected by the mechanical procedure of smear preparation and so are best suited for evaluation. $Fig. 5 (c)-(d)$ shows the phase map of the RBCs.$Fig.6 (a)$ shows the phase of single RBCs form a healthy sample and $Fig.6 (b)$ shows the phase for an unhealthy sample. \\

\begin{figure}[ht!]
\centering\includegraphics[width=8cm,height=4cm]{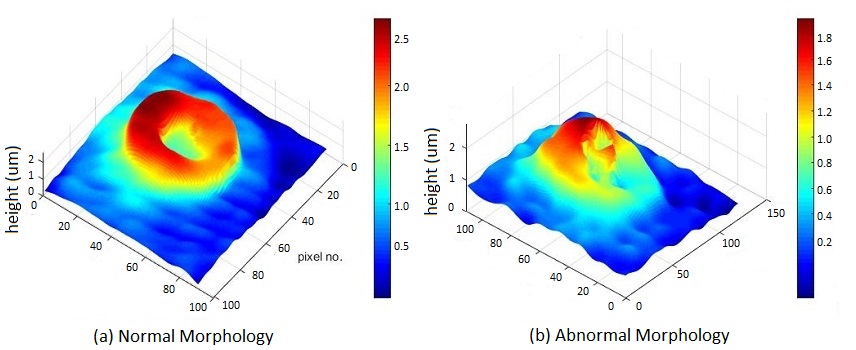}
\caption{(a) Shows the phase for a healthy RBC, (b) Shows the phase for an unhealthy RBC.}
\end{figure}

\section{Conclusion:}
A laboratory breadboard set up has been realized for the phase imaging of the cellular objects present in the blood sample. It is a compact setup that uses only few optical components as compared to two beam interference holographic microscope. The components used in the present set up are a laser diode, microscopic objective, shear plate and a CMOS digital array. The holograms were recorded using the altered and unaltered part of the same FOV of the sample slide by specially focusing the peripheral part of the blood smear to avoid the duplicate (overlap) image problem. The reflected beams from the two  surfaces of the shear plate had almost same intensity (4\% approx for both interfaces), leading to the best contrast fringes. This helps in retrieving the object phase with good accuracy. The setup gives high temporal stability and phase results comparable to those of  the two beam interference holographic microscope. This works well for the sparse samples as it avoids overlapping. However, the background in such a case may  not be uniform leading to  small distortions in the object phase. The FOV can also be increased by filtering one of the beams. This type of geometry requires an additional lens and a pinhole of very small diameter. The method may lead to the loss of phase information due to the finite extent of the pin hole. Because of the common path geometry, pin hole may also block the high frequencies components of the object beam. The proposed geometry and methods are free from such issues, however the overall throughput is relatively small, which can be increased by using a smear prepared with extended boundaries. This will involve preparation of multiple smears on the same slide. This is being worked upon and will be presented in the next manuscript.
\bibliographystyle{plain}
\bibliography{bibliography.bib}

\end{document}